# Asteroids and the James Webb Space Telescope

Andrew S. Rivkin[1], Franck Marchis[2], John A. Stansberry[3], Driss Takir[4], Cristina Thomas[5,6,7] and the JWST Asteroids Focus Group

1. andy.rivkin@jhuapl.edu, Johns Hopkins University Applied Physics Laboratory, 11101 Johns Hopkins Rd. Laurel MD 20723
2. fmarchis@seti.org, SETI Institute, 189 Bernardo Av, Mountain View CA 94043, USA
3. jstans@stsci.edu, Space Telescope Science Institute, 3700 San Martin Drive, Baltimore MD, 21218
4. dtakir@usgs.gov, Astrogeology Science Center, United States Geological Survey, 2255 N. Gemini Dr., Flagstaff AZ, 86001, USA
5. cristina.a.thomas@nasa.gov, NASA Goddard Space Flight Center, 8800 Greenbelt, Rd., Greenbelt, MD 20771
6. NASA Postdoctoral Program, Oak Ridge Associated Universities, PO Box 117, MS 36, Oak Ridge, TN 37831
7. Planetary Science Institute, 1700 East Fort Lowell, Suite 106, Tucson, AZ 85719.

*Abstract:* The James Webb Space Telescope (JWST) provides the opportunity for ground-breaking observations of asteroids. It covers wavelength regions that are unavailable from the ground, and does so with unprecedented sensitivity. The main-belt and Trojan asteroids are all observable at some point in the JWST lifetime. We present an overview of the capabilities for JWST and how they apply to the asteroids as well as some short science cases that take advantage of these capabilities.

*Background:* While asteroids were once derided by frustrated astrophysicists as "vermin of the skies", we have come to realize they are invaluable tracers of history long erased from larger bodies. The non-gravitational processes they experience can involve forces so small that they cannot be measured on more massive bodies. Their present-day orbits contain echoes of giant planet migration, offering constraints that help us understand not only our solar system but others as well. A handful of small and large asteroids have received spacecraft visitors on flyby or rendezvous missions, with one sample return completed and two more in the works. However spacecraft will never visit the vast majority of asteroids, estimated at roughly 1 million larger than 1 km and millions more smaller than that size. These objects must be observed remotely if the scientific questions they answer are to be addressed. As the flagship space telescope of the coming decade, the James Webb Telescope (JWST) is an obvious place for planetary astronomers to turn.

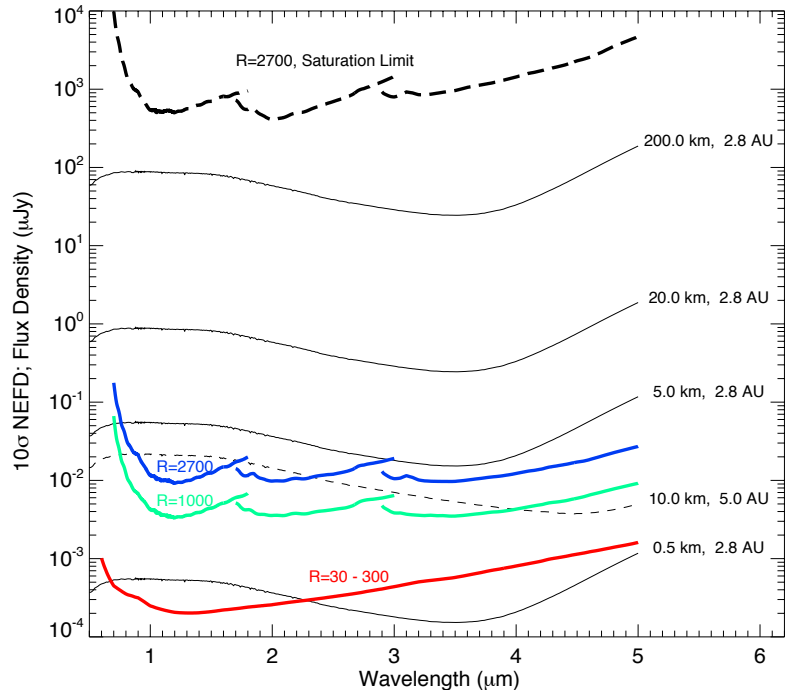

**Figure 1:** Model spectra for 4 main-belt and one Trojan asteroid are compared to NIRSpec sensitivity curves and saturation limits. Asteroids are assumed to have 5% geometric albedo over the entire wavelength range shown. Thin solid lines are for the main-belt asteroids, the thin dashed line is for the Trojan asteroid (assumed to be at 5 AU). Thick colored lines give the sensitivity (noise-equivalent flux density, NEFD) for the three resolving powers available with NIRSpec. The thick dashed line gives the point-source saturation limit for the highest resolution mode. Main-belt asteroids larger than about 200 km will be resolved in this wavelength range so it should be possible to observe nearly any MBA using NIRSpec without saturating the detectors.

The small bodies are unusual targets for JWST compared to most solar system objects. Their sizes are often uncertain and typical lightcurve effects can change brightnesses by a few tenths of magnitudes in an hour or so. However, their centrality to important questions in planetary science was recognized in the Planetary Science Decadal Survey, and JWST observations have the potential to revolutionize their study.

The small bodies have been subject to several classification schemes relating to their physical appearance, orbit, formation location, composition, and other factors. There is evidence that groups of small bodies now in overlapping parts of the solar system may have formed at quite different solar distances (Walsh et al. 2012), and that small body populations that formed together may be scattered over the length of the solar system (Gomes et al. 2005, Morbidelli et al. 2005). In this white paper we focus on the group of small bodies with semi-major axes from roughly 2-5.2 AU, taking in the main asteroid belt and the Hilda and Trojan groups, with a few nuances: regardless of semi-major axis we exclude the near-Earth objects (defined as having perihelia < 1.3 AU), which have several unique challenges and are the subject of the white paper of Thomas et al., and we also include the irregular satellites of Jupiter in this paper under the heading of the Trojan asteroids. Finally, although formally classified by the International Astronomical Union as a "dwarf planet" rather than an asteroid, we include Ceres.

Asteroid science used the Hubble Space Telescope (HST) primarily for improved spatial resolution, as with observations of Vesta (Zellner et al. 1997, Thomas et al. 1997, Binzel et al. 1997), Ceres (Thomas et al 2005), and Pallas (Schmidt et al 2008). Deconvolution of Wide Field/Planetary Camera (WF/PC) images of even smaller asteroids were used to determine diameters, albedos, and lower limits to axial ratios (Storrs et al. 2005, among others). Observations of "main belt comets" (MBC, also called "active asteroids") also take advantage of the improved spatial resolution, as well as increased sensitivity of HST, though the MBCs were first detected using smaller, ground-based telescopes. JWST will have a much larger primary mirror than HST (25 m² vs 4.5 m²), providing over 5 times more light-gathering power. HST's instruments do not currently operate beyond 1.7 μm (Wide Field Camera 3 – WFC3)) and never operated beyond 2.45 μm (Near Infrared Camera and Multi Object Spectrometer – NICMOS)), while JWST's extended wavelength range allows measurements of hydroxyl and organic fundamental absorptions in the near-IR, and thermal emission in the mid-IR. Furthermore, JWST operates at L2, much further from Earth than HST, enhancing temperature stability and providing longer possible periods of continuous observation. The telescope is optimized for infrared optics, including a cold primary mirror and subsequent optics, and all instruments are cryogenic.

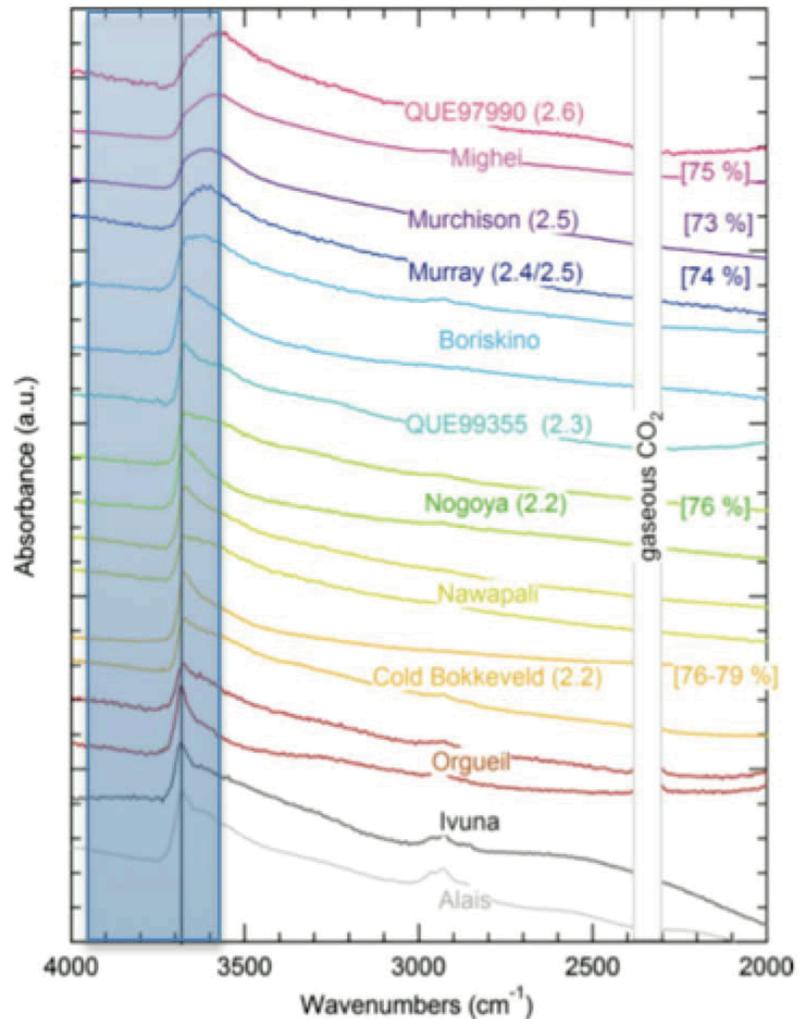

**Figure 2:** Adapted from Beck et al. (2010). The band minimum for carbonaceous chondrites (vertical line), due to hydroxyl in silicates, systematically changes wavelength with metamorphic grade. The variation is entirely in a wavelength region blocked by the Earth's atmosphere (shaded region), making JWST a unique facility to study this variation.

These features mean that JWST should be in general an excellent platform for observations of asteroids. With this in mind, the benefits of JWST to asteroid studies can be effectively divided into three categories: Improved spatial resolution, better S/N on small targets, and ability to reach wavelengths unreachable from ground-based telescopes. Asteroid-centered projects will most likely make use of JWST for one or more of these reasons:

*Improved Spatial Resolution:* The PSF of the NIRCam imager is 64 mas at 2 µm, which corresponds to roughly 80 km in the middle of the asteroid belt (2.8 AU) at opposition, and 45 km at the inner edge of the asteroid belt (2 AU) at opposition. Beyond Vesta and Ceres, both spacecraft targets, there are roughly 25 main-belt asteroids larger than 200 km in diameter, for which shape modeling and compositional mapping could be undertaken. Orbital eccentricity is sufficient to bring perihelion of 8 of the 15 outer belt asteroids 200 km and larger within 2.8 AU, with another 3 reaching 2.9 AU .

*Wider Wavelength Coverage:* Observations in several important wavelength regions are hampered by poor transmission through the Earth's atmosphere. For instance, the band center for hydroxyl in clays gives a good indication of the amount of alteration experienced by carbonaceous chondrites but the entire 2.5-2.8 µm spectral region covering the range of band centers is unobservable from the Earth (Figure 2, Beck et al. 2010, Rivkin et al. 2002). HST observations were not possible at these wavelengths either, due to lack of appropriate instrumentation. At longer wavelengths, the L and M band regions (~3-5 µm) are also both very challenging from the ground and essentially unexplored for asteroids from spacecraft: Spitzer and the Wide-field Infrared Survey Explorer (WISE) only had photometric capabilities at those wavelengths. The Japanese Akari spacecraft had some spectroscopic capability, but made limited observations of asteroids. These wavelengths include absorptions due to organic materials as well as Si-H (a possible product of space weathering). JWST observations provide the opportunity to obtain data in these regions, opening up new means of quantitatively measuring mineralogies, characterizing band shapes and centers, and determining histories. The wider wavelength coverage available from JWST is a major driver for the example projects discussed below.

*Improved Sensitivity:* Not only does poor transmission block some wavelengths from measurement from Earth, but it also drastically affects the limiting magnitude of objects. In the 3-µm region, diagnostic for water, hydroxyl, and organic material, targets are typically limited to V~14 at the NASA Infrared Telescope Facility (IRTF), corresponding to ~50 km in the mid-belt for carbonaceous bodies. Even with the Keck telescope, 3-µm observations on objects fainter than V~15.5 are a challenge (Brown, pers comm.) and data quality is marginal. This is the case even though objects of V~18 are commonly observed with the IRTF at 2.5 µm, where transmission is much higher. Figure 1 shows that objects 0.5 km in size and 5% albedo at 2.5 AU can be observed with S/N approaching 10 at 3 µm in spectroscopic mode with NIRSpec in 1000 seconds. This corresponds to objects with absolute magnitude of 20.3 and an apparent magnitude of 23.1. We note that *every* numbered asteroid (that is to say, every asteroid with a secure orbit) reaches apparent magnitude brighter than this. The equivalent absolute magnitude for Trojan asteroid sensitivity is 15.3, again fainter than any numbered object in that part of the solar system.

This improved sensitivity with NIRSpec on JWST makes discovery and secure orbit determination the limiting factor for observations from the main belt through the Trojan clouds, along with most known irregular satellites of Jupiter as practically every object will be observable with S/N > 10 in 1000 seconds across the entire 0.6-5 µm range with the R=100 prism. Because asteroidal spectral features are typically broad, solid-state absorptions, higher spectral resolutions are rarely required. Spectrophotometric observations can be made even more quickly and/or at higher S/N.

This situation will not change significantly with future surveys. The Large Synoptic Survey Telescope (LSST) is expected to detect 90% of main belt asteroids to H=20 and 50% of MBAs to H=20.7, with 90% of Trojans to H=17.5 and 50% of Trojans to H=17.8 detected (LSST, 2009). Objects of these sizes are still within a factor of a few of the sensitivity limits mentioned above for 1000 second integrations, and are all easily above the sensitivity limits for spectrophotometry with the NIRCam imager in the 1-5 µm region. As a result almost any arbitrarily-chosen known asteroid in the main belt or beyond can be observed by JWST, although due to solar elongation constraints they cannot be observed at any arbitrarily-chosen time.

The sensitivity of the Mid-Infrared Instrument (MIRI) will allow compositional studies of much smaller objects than currently available. Spitzer studies of asteroids showed some with surprising emissivity patterns interpreted as due to "fairy castle" structure in the regolith or a high fraction of infrared-transparent materials (Vernazza et al. 2012, Yang et al. 2013). Observations of smaller targets will allow the degeneracy to be resolved. In addition, asteroids typically have emissivity spectra close to black bodies, with compositional effects playing only a small role. Improved sensitivity will increase the number of targets for which these compositional effects can be seen.

*Improved contrast:* Because of its larger aperture and the stability of its point spread function, JWST instrumentation will allow observations of the close vicinity of asteroids with unprecedented sensitivity. Current extreme adaptive optics (AO) systems are capable of detecting (at 5-sigma) features with δmag ~ 7-8 at 0.4" from a bright target (typically V<10.5). For fainter targets, the quality of the correction available with AO systems quickly degrades. In comparison, NIRCAM with coronagraphic mask should offer a 5-sigma detection with δmag ~10-12 at that same angular distance (Greene et al. 2010). This gain will also be independent of the brightness of target, which is not true of AO systems. The Near-InfraRed Imager and Slitless Spectrograph (NIRISS) will have an aperture masking interferometry (AMI) mode capable of reaching 75 mas resolution and contrast sensitivity of ~10 magnitudes at 4.6 µm through the use of non-redundant aperture masking. This contrast sensitivity is more than, and resolution is similar to, that needed to separate Ida from Dactyl (6.7 magnitudes and ~70 mas, respectively), and will be able to individually observe the Ida/Dactyl-like systems in the inner part of the asteroid belt ($a \lesssim 2.5$ AU) High-quality spectral observations of the innermost satellites of Jupiter, point sources 9-14 magnitudes fainter than Jupiter's surface brightness with angular separations of only 10-

30" from Jupiter's limb, will also be enabled by JWST's improved ability to observe faint objects close to bright ones.

*Examples of projects taking advantage of multiple aspects of JWST:* As noted, we anticipate successful asteroid projects will utilize several of the advantages mentioned above. For instance,

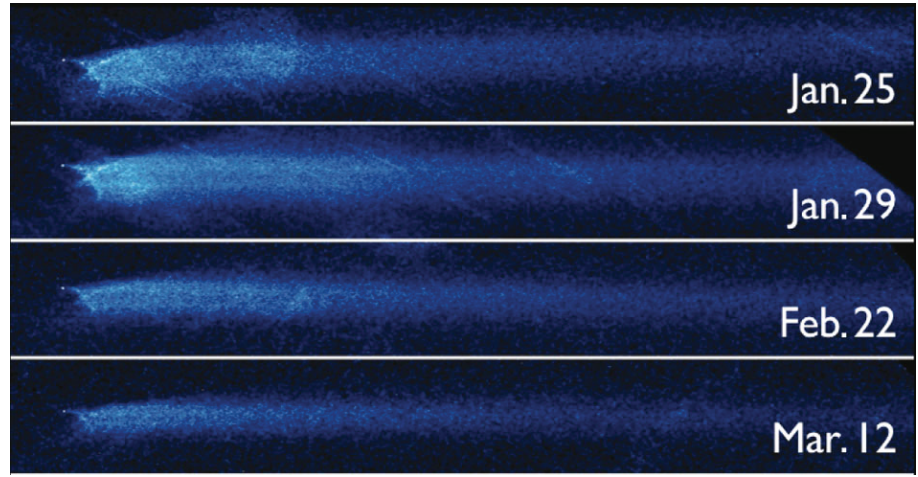

**Figure 3:** The "main belt comets" (MBCs), like P/2010 A2 (figure from Jewitt et al. 2010), have been very productively observed with HST. The visible-near IR-mid IR observations available with JWST will allow us to determine which of these objects are icy or rocky, measure particle sizes, and watch the evolution of its properties with time.

10 Hygeia (430 km diameter) reaches 2.77 AU and subtends 0.34" at perihelion, and 3 Juno (233 km diameter) reaches perihelion inside of 2 AU, subtending 0.32". Both of these values are very close to the angular size of Pallas when it was observed by Schmidt et al. (2009), shown in Figure 4. NIRSpec will allow image cubes to be obtained of these objects, rather than simple spectrophotometry—this is also shown schematically in Figure 4, where every pixel is associated with a visible-near IR spectrum. The ability to observe over a continuous period, unlike HST, enables more straightforward coverage over an entire rotation (compared to piecing together observations from different HST orbits).

A second example project would be studying "main belt comets" (MBCs: Figure 3). These objects have cometary tails and comae but are in main-belt orbits. Some of these objects are thought to be shedding material after impact events or spin up due to the Yarkovsky-O'Keefe-Radzievskii-Paddack (YORP) effect, but others appear to have subsurface volatiles driving their activity. Volatiles have never been detected on these bodies, but because almost all of them are 5 km in size or smaller, constraints are quite loose. Observations with NIRCAM and MIRI would easily establish whether ice is exposed on their surfaces as well as provide firmer sizes and albedos, and observations of MBC tails across the NIRCAM and MIRI wavelengths would enable more precise measurements of particle sizes. The rate of MBC discoveries suggests there will be several available during the lifetime of JWST. Searches for gas in MBC comae (as opposed to dust) have placed water outgassing rates of $\sim 10^{26}$ s$^{-1}$ for several objects (Jewitt 2012). Kelley et al. (this volume) calculated S/N for JWST observations of comets, and note that water outgassing at this rate is detectable to $\sim 4.5$ AU. We can therefore expect to improve upon existing constraints throughout the main asteroid belt.

A third potential project is observations of asteroid collisional family members. For decades we have expected to see variation in the spectra of family members corresponding to, for instance, crust/mantle/core fragments of differentiated bodies. However, observations in the 0.8-2.5 μm region have found family members to be remarkably similar to one another even when parent bodies are suspected to be differentiated (Bus and Binzel 2002, Bus 2014).

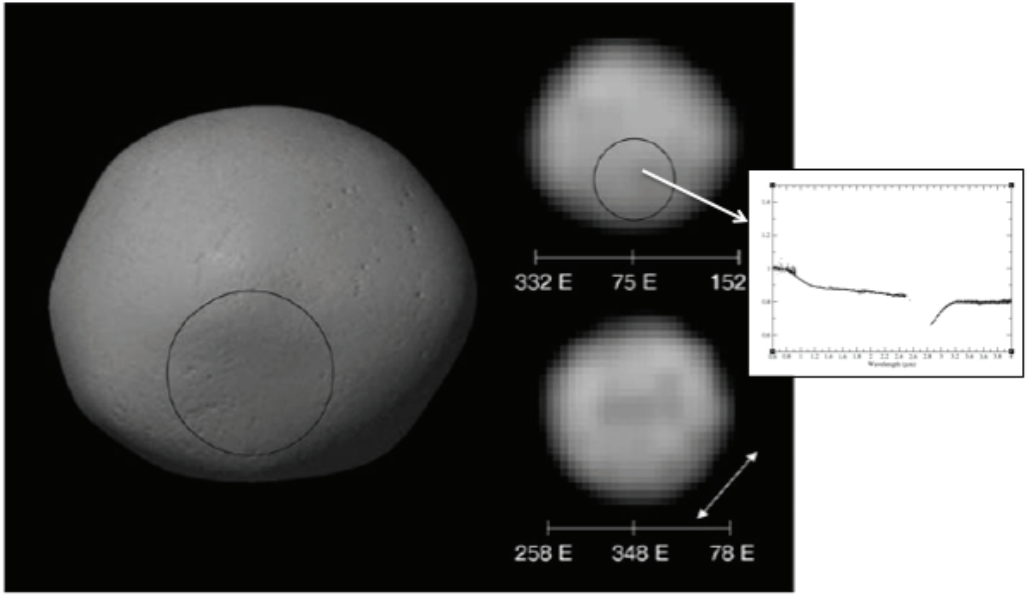

**Figure 4:** Schmidt et al.'s HST observations of Pallas serve as examples of the kind of spatial resolution possible for several large asteroids with JWST. However, unlike HST, the NIRSPEC instrument provides spectra at each pixel (schematically represented by the arrow and reflectance spectrum), allowing compositional variation to be more readily and definitively established when present. Figure from Schmidt et al. (2008)

This has been interpreted as showing that the collisions creating families disrupt bodies into very small pieces, which then reaccrete. Using JWST, we can reach smaller targets in collisional families and potentially determine the characteristic size of collisional fragments from the size at which spectral diversity appears. Current family studies are typically conducted on 3-4 m class telescopes and limited to objects of apparent 19th magnitude or brighter, corresponding to absolute magnitudes of H ~ 15-16.5 for a range of semi major axes from 2.3-3 AU, or diameters of roughly 1-6 km for that range of distances and a range of albedos from 5-30%. As noted above and in Figure 1, high-quality NIRSPEC observations can be made for much smaller objects, and NIRCAM observations can probe targets perhaps 10% the size of the current limit.

*When does MIRI saturate on asteroids?*: The extreme sensitivity of MIRI has the consequence that it saturates for surprisingly bright targets. Ceres saturates every filter and grating other than the shortest-wavelength half of the shortest-wavelength grating. Because saturation is related to surface brightness, smaller objects with Ceres' temperature will still saturate MIRI as long as they are larger than 0.11" in angular diameter, the size of a MIRI pixel. Angular sizes below 0.11" may or may not saturate MIRI depending on a combination of the fraction of the pixel filled and temperature. Table 1 shows estimates of the temperatures that are expected at various important distances for asteroids: the inner edge of the asteroid belt at 2.0 AU, the middle of the belt near where Ceres and Pallas orbit at 2.8 AU, the Cybele region at 3.4 AU just beyond what is typically considered the outer edge of the main belt, and 5.2 AU for Trojan asteroids and the irregular satellites of Jupiter.

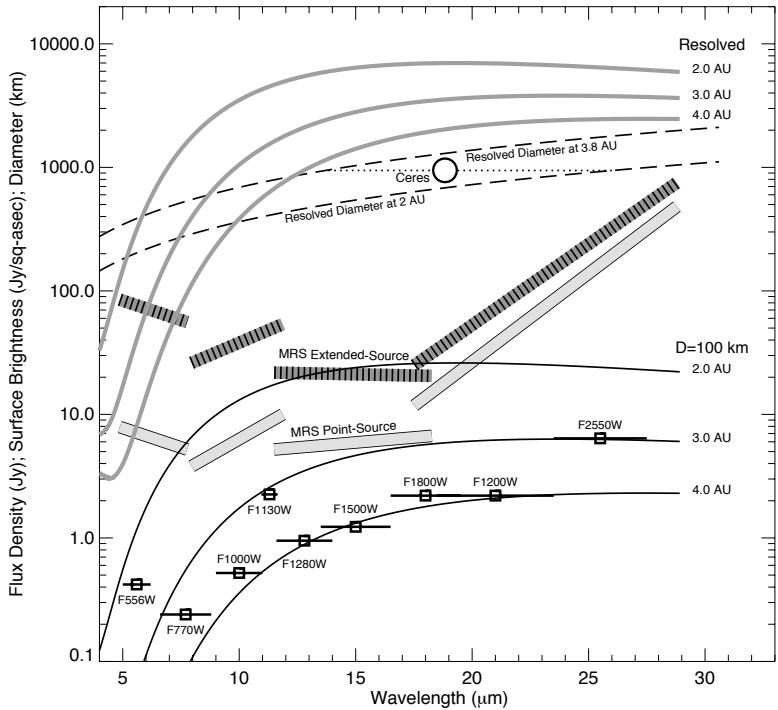

**Figure 5:** Model spectra for main-belt asteroids are compared to the saturation limits of MIRI. Two cases are considered: asteroids that will be un-resolved point sources, and those that are large enough to be resolved. The spectra of un-resolved asteroids with diameters of 100 km at heliocentric and JWST distances of 2, 3 and 4 AU are shown as solid black lines. Solid gray lines give the surface-brightness emission expected from resolved asteroids at those same distances. These models assume geometric albedo of 0.15, phase-integral of 0.39, and beaming parameter of 0.9. Long-dashed black lines give the diameter of objects that will be resolved at distances of 2.0 AU and 3.8 AU (inner and outer belt) as a function of wavelength. Ceres (diameter of 943 km at a distance of 2.77 AU) is resolvable at wavelengths below 18.4 um. Other symbols give the MIRI saturation limits. Black bars with square symbols and labels for the MIRI filters give the point-source imaging saturation limits in 64x64 pixel subarrays. Wide, light-gray bars give the Medium Resolution Spectrometer (MRS) point-source saturation limits. Dark-gray bars with black stripes give the MRS saturation limits for extended sources (e.g. resolved asteroids). Nearly all resolved asteroids will saturate MIRI in all modes, MRS channel 1 (5-8 um) being the possible exception. Asteroids as large as 100 km diameter should be observable via imaging or the MRS, depending on their distance from the Sun and JWST. Saturation limits shown here follow Glasse et al. (2015).

These temperatures assume a spherical body and are calculated for a position 45° from the subsolar point using a geometric albedo of 0.2, G=0.15 (resulting in a Bond Albedo A of 0.08), emissivity (ε) of 0.9, and a beaming parameter (η) of 1. The choice of 45° from the subsolar point represents a balance between the contribution of the hottest part of the surface and the contribution from a larger surface area. The subsolar point will have a temperature 8% higher than the locations 45° from the subsolar point. The subsolar temperature ($T_{ss}$) is calculated using the well-known equation

$$T_{ss} = ((1-A)\, S\, /(\eta\, \varepsilon\, \sigma))^{1/4},$$

where S is the solar constant at the distance of interest and σ is the Stefan-Boltzmann constant (Lebofsky et al, 1986, for example).

Figure 5 overlays the surface brightness of targets of these various temperatures for objects large enough to fill a MIRI pixel. The 3rd-6th columns of Table 1 show the sizes at which MIRI pixels are just filled (at various asteroid distances, assuming JWST is at 1.0 AU and observing at opposition), and the sizes at which various observing modes become possible as a function of distance.

The largest Trojan asteroid (624 Hektor) has a diameter of ~225 km. All Trojans are capable of being observed by MIRI in the 10-µm gratings, though 20-µm spectroscopy cannot be done until smaller sizes are reached. The largest few irregular satellites of Jupiter are also too large for 20-μm spectroscopy or 15-μm imaging, though over 50 of the irregular satellites will not saturate MIRI.

| Distance (AU) | T (K) | Pixel fill (km) | 10-μm Size (km) | 20-μm Size (km) | 15-μm Size (km) |
|---|---|---|---|---|---|
| 2.0 AU | 260 | ~80 | ~15 | ~6 | ~3 |
| 2.8 AU | 220 | ~150 | ~40 | ~12 | ~8 |
| 3.4 AU | 200 | ~200 | ~70 | ~20 | ~12 |
| 5.2 AU | 150 | ~330* | ~330* | ~60 | ~55 |

**Table 1: Observable asteroid sizes for MIRI. The third column is the size at which a MIRI pixel is filled. The three rightmost columns are the maximum size for observations without saturation for spectroscopy in the 10- and 20-μm regions and imaging at 15 μm. *The largest Trojan asteroids and Jovian irregular satellites are smaller than 330 km, so all are observable in these modes.**

*Conclusions:* Because of its powerful capabilities, JWST provides unique opportunities for asteroid science. It is likely to be used in solving several important small bodies questions during its lifetime, though naturally it is not the ideal facility for all observations and some observations will be impossible due to saturation. Nevertheless, the small bodies community should take advantage of JWST to the extent possible to obtain data impossible to acquire using other methods.

*Acknowledgments:* ASR would like to acknowledge support from NASA Planetary Astronomy Grant NNX14AJ39G and NSF Planetary Astronomy Award 1313144. The authors would like to thanks members of the JWST Project at NASA Goddard and staff members at STScI for information and review of this manuscript.